\documentclass[aip,reprint,showpacs,superscriptaddress]{revtex4-1} 

\usepackage{amsmath,amssymb,graphicx}
\usepackage[english]{babel}
\usepackage{adjustbox}

\usepackage{natbib}
\usepackage[colorlinks=true,linkcolor=blue]{hyperref}%

\usepackage{color}

\DeclareRobustCommand{\_}[1]{\sb{\mathrm{#1}}}

\begin{document}

\title{Thickness dependent enhancement of the polar Kerr rotation in Co magnetoplasmonic nanostructures}

\author{Richard M. Rowan-Robinson}
\email{richard.rowan-robinson@physics.uu.se}
\thanks{Author to whom correspondence should be addressed.}
\affiliation{Department of Physics and Astronomy, Uppsala University, Box 516, SE-751 20 Uppsala Sweden}

\author{Emil Melander}
\affiliation{Department of Physics and Astronomy, Uppsala University, Box 516, SE-751 20 Uppsala Sweden}

\author{Ioan-Augustin Chioar}
\affiliation{Department of Physics and Astronomy, Uppsala University, Box 516, SE-751 20 Uppsala Sweden}

\author{Blanca Caballero}
\affiliation{IMM-Instituto de Microelectronica de Madrid (CNM-CSIC), Isaac Newton 8, PTM, Tres Cantos, E-28760 Madrid, Spain}

\author{Antonio Garc\'{\i}a-Mart\'{\i}n}
\affiliation{IMM-Instituto de Microelectronica de Madrid (CNM-CSIC), Isaac Newton 8, PTM, Tres Cantos, E-28760 Madrid, Spain}

\author{Evangelos Th. Papaioannou}
\affiliation{Fachbereich Physik and Forschungszentrum OPTIMAS, Technische Universit\"{a}t Kaiserslautern, 67663 Kaiserslautern, Germany}

\author{Vassilios Kapaklis}
\email{vassilios.kapaklis@physics.uu.se}
\thanks{Author to whom correspondence should be addressed.}
\affiliation{Department of Physics and Astronomy, Uppsala University, Box 516, SE-751 20 Uppsala Sweden}

\begin{abstract}

Large surface plasmon polariton assisted enhancement of the magneto-optical activity has been observed in the past, through spectral measurements of the polar Kerr rotation in Co hexagonal antidot arrays. Here, we report a strong thickness dependence, which is unexpected given that the Kerr effect is considered a surface sensitive phenomena. The maximum Kerr rotation was found to be -0.66 degrees for a 100~nm thick sample. This thickness is far above the typical optical penetration depth of a continuous Co film, demonstrating that in the presence of plasmons the critical lengthscales are dramatically altered, and in this case extended. We therefore establish that the plasmon enhanced Kerr effect does not only depend on the in-plane structuring of the sample, but also on the out-of-plane geometrical parameters, which is an important consideration in magnetoplasmonic device design. 

\end{abstract}


\maketitle

\section{Introduction}

Plasmonics allow for the confinement of light on length-scales smaller than the incident wavelength, leading to dramatic enhancements of the electric field within the confining material. Magnetoplasmonics marries ferromagnetism with plasmonics and aims to exploit this field-enhancement in order to produce active optical devices which are tunable, by an external magnetic field~\cite{bossini2016,floess2018,maksymov2016,armelles2013,armelles2014}. The Kerr effect is a well studied example by which magnetism can be used to alter the polarization state of light. However, in pure ferromagnetic thin films the magnitude of the effect is comparatively small. Nevertheless, it has been shown that it can be dramatically enhanced when magnetic materials are combined with plasmonic materials. This enhancement of the  magneto-optical (M-O) effects, especially that of the  polar and transversal Kerr effects, has been reported for both types of plasmonic excitations: localized surface plasmons (LSPs)~\cite{maccaferri2016,armelles2016,bonanni2011,kataja2016,sepulveda2010,maccaferri2013,lodewijks2014}- and for propagating surface plasmon polaritons (SPPs)~\cite{ctistis2009,maccaferri2015,rollinger2016,martin-becerra2010,temnov2010,belotelov2007,belotelov2011}. Examples include hybrid nanostructures of noble metal/ferromagnetic structures such as Co/Au~\cite{papaioannou2014}, YIG/Au~\cite{belotelov2011}, and Au/Co/Au trilayers \cite{safarov1994,gonzalez-diaz2007,martin-becerra2010,hamidi2018}, as well as in patterned pure magnetic films~\cite{chen2016,melander2012,papaioannou2010,papaioannou2011,chetvertukhin2013,papaioannou2017,zhang2017,fang2015,torrado2010}.

In all cases, the mechanism for the increased M-O response can be attributed to the enhanced and localized electric field provided by the plasmon excitation, as very recently shown by correlating the near-and far-field optical and magneto-optical responses~\cite{rollinger2016}. In this study we investigate the contribution of finite size effects from varying the thickness of the ferromagnetic layer, which, despite the numerous studies on materials combinations and pattern geometry, has received little attention. Previously, \citet{gonzalez-diaz2007} used the Kretschmann configuration with Au/Co/Au trilayers to investigate the enhancement of the transverse magneto-optical Kerr effect signal, for different Co thicknesses, as a result of propagating SPP modes at the Au/air interface. They found an optimum Co thickness of 6~nm, where the redistribution of the electromagnetic field in the magnetic layer due to the SPP excitation at the Au/air interface is the strongest. However, since they were investigating a hybrid structure, whereby the SPP was excited at the Au interface, their study was limited to a maximum Co thickness of just 9~nm. This is due to the relatively short decay length of the evanescent wave from the SPP into the metal along the interface normal direction. Typically, the penetration depth ($\delta$) defines the thickness range over which the enhancement can occur, which for Co is $\approx$~13~nm~\cite{temnov2010} within the visible range of wavelengths. However, in patterned hole array structures (or so-called antidot structures) where SPPs can be excited at normal incidence, $\delta$ can be dramatically enhanced, yielding an effective $\delta$ which can extend above 100~nm~\cite{degiron2002}. The presence of holes in an otherwise continuous film allows for SPPs to couple to both sides of the film, resulting in ``extraordinary optical transmission''~\cite{ebbesen1998}. Wood's anomalies also exist as a result of SPPs, whereby one of the diffracted orders is transmitted tangentially to the array and along the film surface~\cite{ghaemi1998}.

\begin{figure*}
\includegraphics[width = 0.9\textwidth]{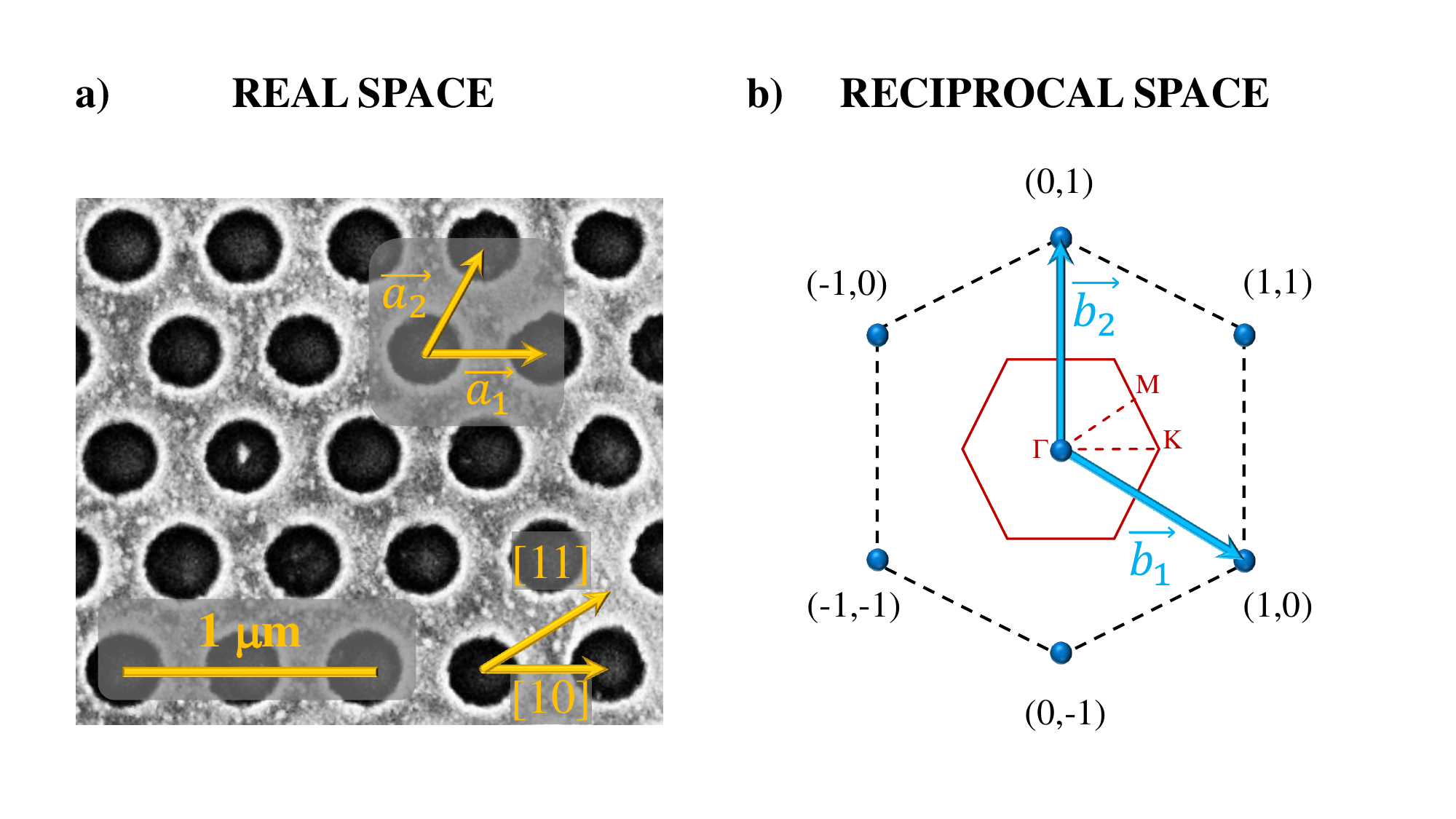}
\caption{\label{fig1}a) SEM image of a typical sample. The real-space unit vectors are defined on the hexagonal hole arrangement by the vectors $\vec{a}_{1}$ and $\vec{a}_{2}$. The principal directions for nearest-neighbour and next nearest-neighbour are therefore [10] and [11] respectively. b) The corresponding reciprocal lattice with reciprocal-space unit vectors $\vec{b}_{1}$ and $\vec{b}_{2}$. The first Brillouin zone is shown in red. All measurements were made along the \textbf{$\Gamma$M} direction, which corresponds to the [11] direction in real-space. }
\end{figure*}

Recently \citet{luong2018} completed a detailed study investigating the thickness dependence of the Faraday effect in Co antidot arrays on glass substrates. They found a maximum Faraday rotation for antidot arrays with a thickness of 30~-~50~nm. This was due to resonant excitation of SPPs and/or field-enhancement at Wood's anomalies. In a similar work, \citet{caballero2015} showed that this thickness range can be increased by incorporating thin gold layers. However, even then, when working in the transmission configuration, limitations are applied to the available useful thickness range due to increased absorption in the film~\cite{caballero2015}. It is therefore interesting to consider the thickness dependence when working in reflection, which would not be subject to the same thickness limitations, as it is inherently a surface sensitive probe.

In this communication, we investigate the thickness dependence of the polar Kerr effect in reflection, for a series of Co antidot films with thicknesses in the range 20~-~100~nm. We combine measurements of the spectral optical reflectivity and spectral polar Kerr rotation, together with theoretical simulations in order to understand the thickness dependence of the SPP enhanced M-O activity. The polar Kerr effect is inherently a surface sensitive phenomena and would generally be considered insensitive to thicknesses above the penetration depth for continuous films. However, we reveal a strong dependence of the polar Kerr rotation on the magnetic layer thickness when the effect is coupled to resonant plasmonic excitations. The study is focused on Co films, with film thicknesses ($t$) smaller than the wavelength of the incident light ($\lambda$), but larger than the penetration depth ($\delta$) of the bulk Co metal.

\section{Experimental}
\subsection{Sample fabrication}
Three Co films were patterned by the use of self-organization of colloidal polystyrene beads on Si substrates as shadow masks, see Ref.~\cite{melander2012} for further details.  The final layout of the samples is presented in Fig.~\ref{fig1}~a). The hexagonal hole structure has a periodicity of $a_{0} = 470\pm15$~nm and hole diameters of $d = 260\pm10$~nm.  A 2~nm buffer layer of Ti was initially deposited for better adhesion of the Co onto the Si, and Co layers with different thickness were grown onto this seed layer. To prevent oxidation of the Co surface, a capping layer of 2 nm Au was deposited. All layers were grown with electron beam evaporation. The final structure of the samples is: Si(111)/Ti (2~nm)/Co (X~nm)/Au (2~nm), with X being 20, 60, or 100~nm. A continuous thin film with X = 20~nm was also prepared at the same time, to be used as a reference sample. The hexagonal hole arrangement is described by the real-space unit vectors $\vec{a}_{1}$ and $\vec{a}_{2}$, as shown in Fig.~\ref{fig1}~a). From this basis, the nearest-neighbour and the next nearest-neighbour directions are defined as [10] and [11] respectively. These directions are of particular importance for the activation of SPPs~\cite{papaioannou2011}. The corresponding reciprocal lattice vectors, $\vec{b}_{1}$ and $\vec{b}_{2}$, are shown in Fig.~\ref{fig1}~b). The first Brillouin zone is shown in red. All measurements were performed with the light polarization along the \textbf{$\Gamma$M} direction, which corresponds to the [11] direction in real-space.

\subsection{Experimental methods}
The Kerr rotation is measured in the polar configuration as a function of the incident light wavelength. A mercury lamp source is used to obtain a broadband white light spectrum, which is then monochromatized using a grating monochromator. After the monochromator, the light is sent first through a high-pass filter, to remove any higher harmonics. Following this, it passes through a set of beam-shaping and focusing lenses to maximize the light intensity reaching the sample and through the bore in the electromagnet pole piece. Prior to the sample, the light passes through a fixed initial polarizer, which is used to define the polarization state with respect to the principle axes of the hexagonal antidot array. Inside the electromagnet the light is incident on the sample at an angle of 4 degrees. The maximum magnetic field strength is 1.6 T.

To minimize the noise, the light is modulated with either a chopper (for $\lambda <$ 400 nm) or a Faraday cell (for $\lambda >$ 400 nm). Light which is specularly reflected from the sample passes out through a second bore in the pole piece, and then through a second polarizer before being focused onto the detector.  The modulated signal is measured using either a Si-photodiode detector or a photo-multiplier tube connected to a lock-in amplifier. The second polarizer is automated, such that it rotates until the mimimum detector signal is obtained, corresponding to the crossed-polarized configuration. This experiment configuration reflects a simplified scenario of the generalized magneto-optical ellipsometry method~\cite{berger1997}. The polar Kerr rotation is measured in absolute values (degrees) by measuring the difference between the extinction angles of the second polarizer $\theta\_{ext}$, as  measured after magnetically saturating the sample in the two polar directions. The absolute Kerr rotation is calculated from~~$\theta\_{Kerr}= \frac{1}{2}( \theta\_{ext}(+H) - \theta\_{ext}(-H))$. The same setup is used for the spectral reflectivity measurements, but, for this scenario, the specularly reflected light is guided directly onto the detector and the second polarizer is removed. Reflectivity is measured relative to the intensity of the direct beam. For this work, spectral reflectivity and Kerr rotation were measured for an energy range from 1.5~eV up to ~4~eV ($\lambda = $310~-~820~nm). The polar Kerr rotation spectra was measured with the sample in a saturated magnetic state (applied magnetic field $\mu_0 H = 1.1$ T).

\section{Results}
\subsection{Optical reflectivity}
\begin{figure*}
	\centering
	$\vcenter{\hbox{\adjincludegraphics[width=0.9\columnwidth,trim={0 0 0 0},clip]{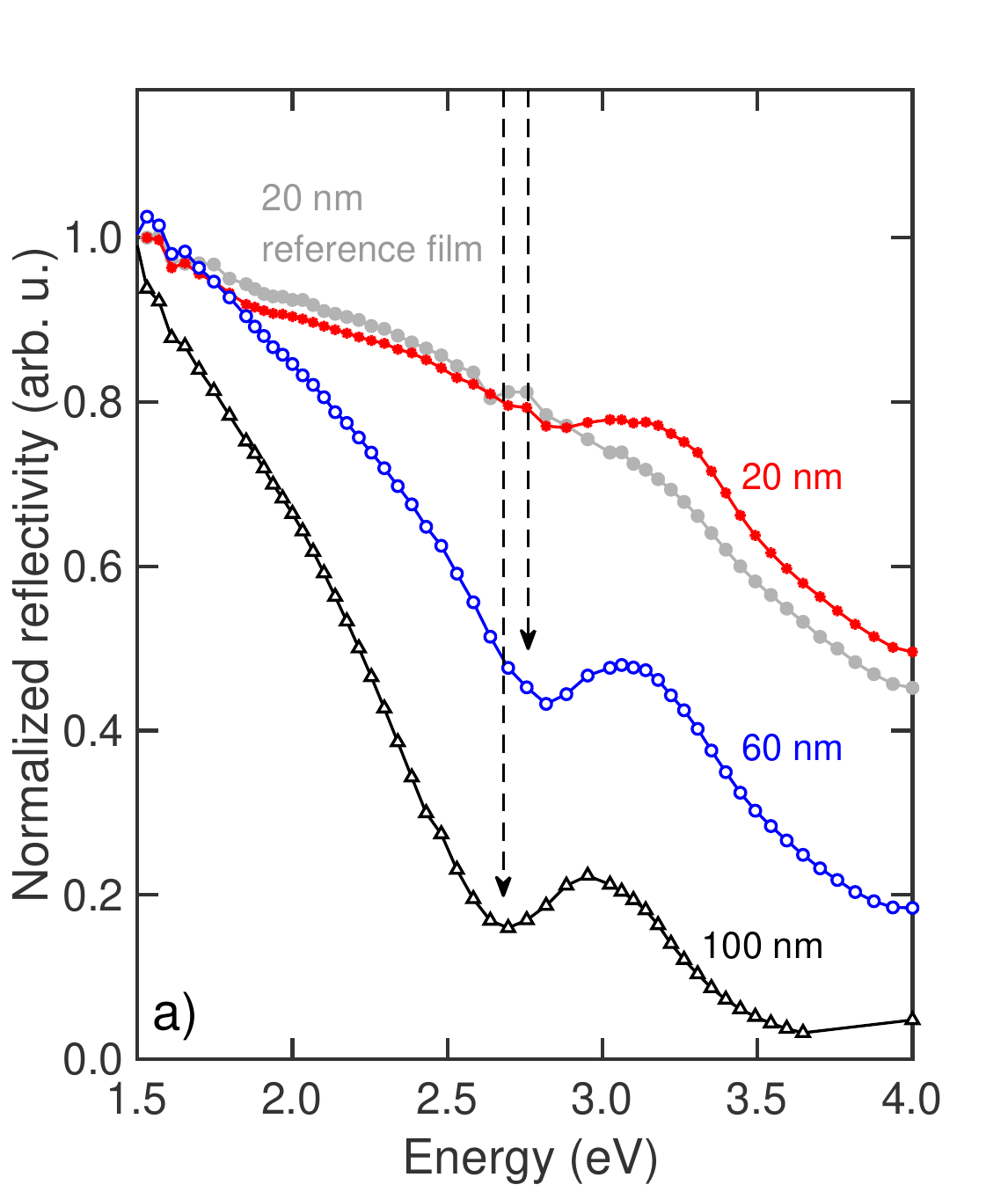}}}$%
	$\vcenter{\hbox{\adjincludegraphics[width=0.9\columnwidth,trim={0 0 0 0},clip]{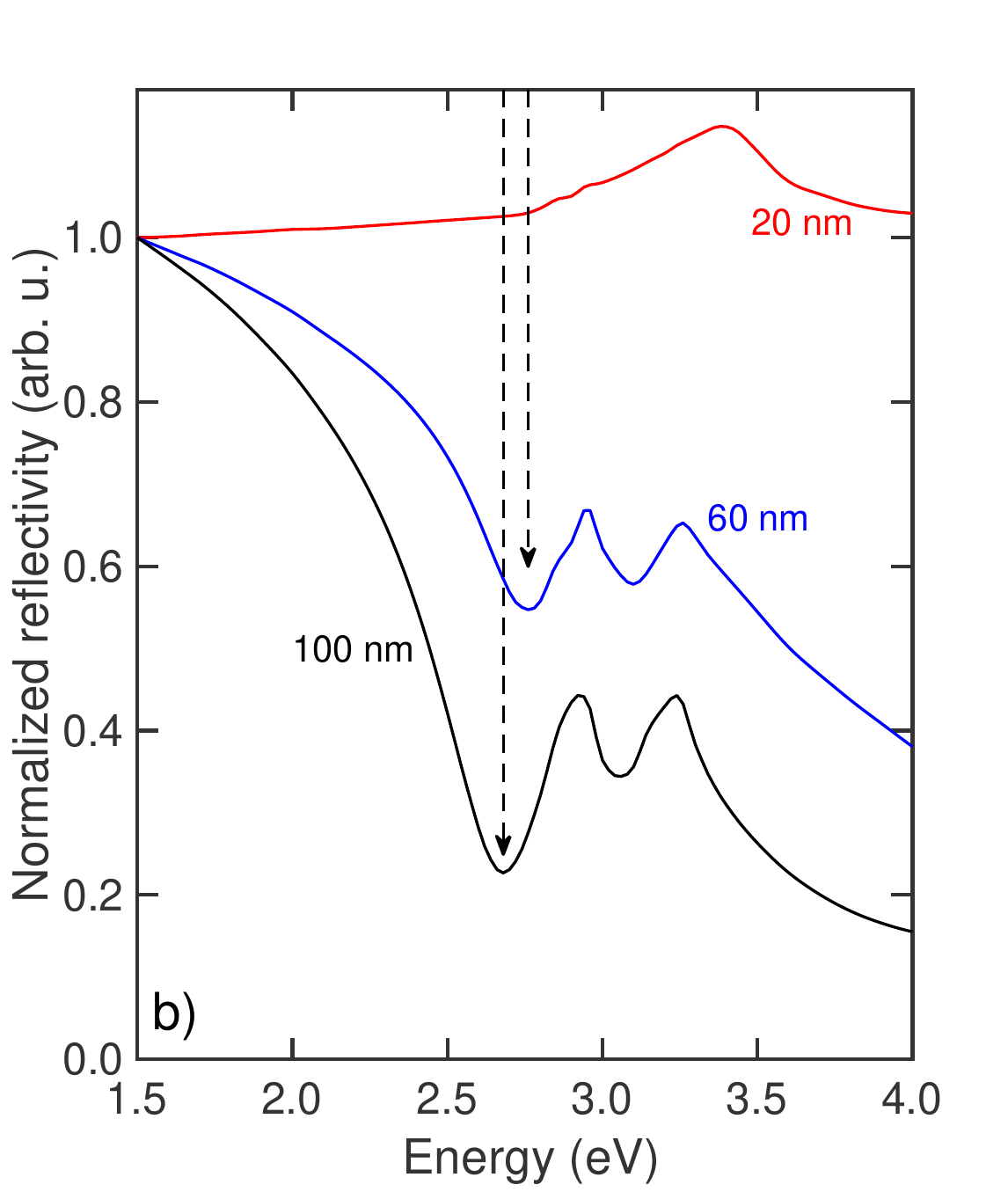}}}$
	\caption{a) Measured specular reflectivity for 100~nm, 60~nm, 20~nm antidot films as well as the 20~nm reference film as function of incidence light energy. b) Calculated reflectivity using the scattering matrix approach for the 20~nm, 60~nm and 100~nm antidot films. In both figures, the data has been normalized to the 1.5 eV value of each dataset to aid the clarity of comparison. The angle of incidence is 4 degrees for both the simulation and experiment. The arrows indicate the predicted energies at which SPPs would be excited at the Co/air interface based on the calculated reflectivity. Reflectivity minima are observed at the predicted energies where SPPs would be excited. }
	\label{fig2}
\end{figure*}

The measured specular reflectivity of the antidot samples is shown in Fig.~\ref{fig2}~a), with the polarization aligned along the next nearest-neighbour [11] direction and is \textit{p}-polarized with respect to the scattering plane defined by the 4 degree angle incident beam and the normal to the sample surface. The measured data has been normalized to the 1.5~eV value. Both the 20~nm continuous film and the 20~nm antidot samples show similar behaviour, with a decrease in the specular reflectivity as the energy increases. For the 60~nm antidot sample, a minimum in reflectivity occurs at $\approx$ 2.81 eV. This feature becomes more pronounced for the 100~nm sample and is redshifted to $\approx$ 2.69 eV. At higher energy, a broad but intense minimum occurs around 3.75~eV and 4.0~eV for the 100~nm and 60~nm samples respectively. However, since the Si substrate becomes highly absorbing for energies above 3.2~eV, these features are treated with a degree of caution. Furthermore, a similar feature is observed in the 20~nm continuous film, thereby suggesting it has no correspondence to the antidot array or any plasmonic origin. We consider that the features at 2.81~eV and 2.69~eV for the 60 nm and 100 nm samples respectively are the result of resonant coupling of the light to the 2D lattice, via SPPs excited at the Co/air interface. Due to the high refractive index of Si, the primary SPP excitations at a Si/Co interface would occur at much lower energies and can not explain the features at 2.81~eV and 2.69~eV. Furthermore, absorption within the Co film severely limits any contribution from the buried Si/Co interface.   

The reflectivity spectrum calculated using the scattering matrix approach~\cite{caballero2012} is shown in Fig.~\ref{fig2}~b). A reduction in reflectivity is observed when SPPs are excited, and the simulation shows good agreement with the experiment as indicated by the vertical arrows depicting the calculated energies of the SPPs associated with the Co/air interface. In particular, the calculated reflectivity spectra reproduces the redshift with increasing thickness, by which the primary Co/air SPP resonance shifts from 2.76~eV to 2.68~eV when increasing the Co thickness from 60 to 100~nm.

To be consistent with the experiment, the theoretical curves were calculated for a 4 degree incidence angle, and consequently an additional feature appears in the calculated reflectivity at $\approx$~3.1 eV. This originates from the lifting of the degeneracy between the plasmonic modes when not operating at normal incidence. In the experiment we do not see these dips in the measured specular reflectivity curves. It is considered this is due to their lower relative intensity in combination with sample geometrical imperfections and the smaller resolving power of our reflectivity measurement. The effect of this is to smear out any plasmonic features as compared to the theoretical curves which assume perfect lattice periodicity and perfectly smooth interfaces.

When discussing the thickness dependence, we consider that with reducing Co thickness the electronic properties may begin to differ from that of the bulk material. It could be expected that by reducing the thickness, the electrical conductivity would decrease, causing the real and imaginary parts of the refractive index to vary from that of bulk. This has been observed for un-patterned Ni and Al films, for which the refractive index exhibits a thickness dependence even above the film percolation thickness, which can be attributed to grain-boundary scattering~\cite{nguyen1993,kamineni2009}. 

In order to try to accommodate for these finite size effects, the optical constants for Co used in the calculation were obtained from experimental measurements of 20~nm Co thin films~\cite{ferreiro-vila2009}. However, this may still be insufficient to account for increased roughness or reduced grain size, which can result as a consequence of the lithographic processing. A reduced conductivity would result in poorer screening of the electric field around the antidot, resulting in a larger effective hole diameter. This in turn would result in broader, lossy and less intense resonances, as is observed in Fig.~\ref{fig2}~a).

\subsection{Plasmon enhanced polar Kerr rotation}
\begin{figure*}
	\centering
	$\vcenter{\hbox{\adjincludegraphics[width=0.9\columnwidth,trim={0 0 0 0},clip]{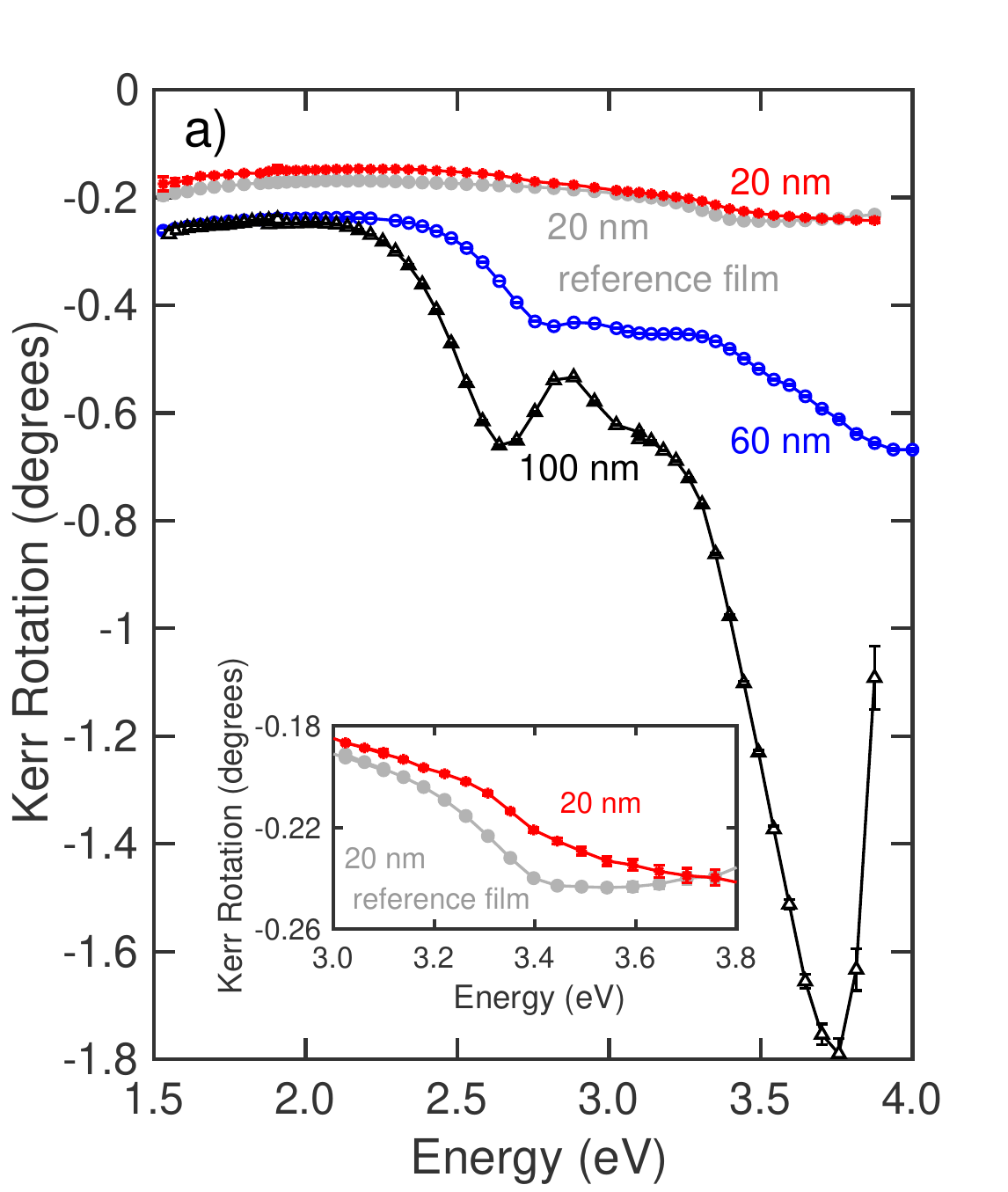}}}$%
	$\vcenter{\hbox{\adjincludegraphics[width=0.9\columnwidth,trim={0 0 0 0},clip]{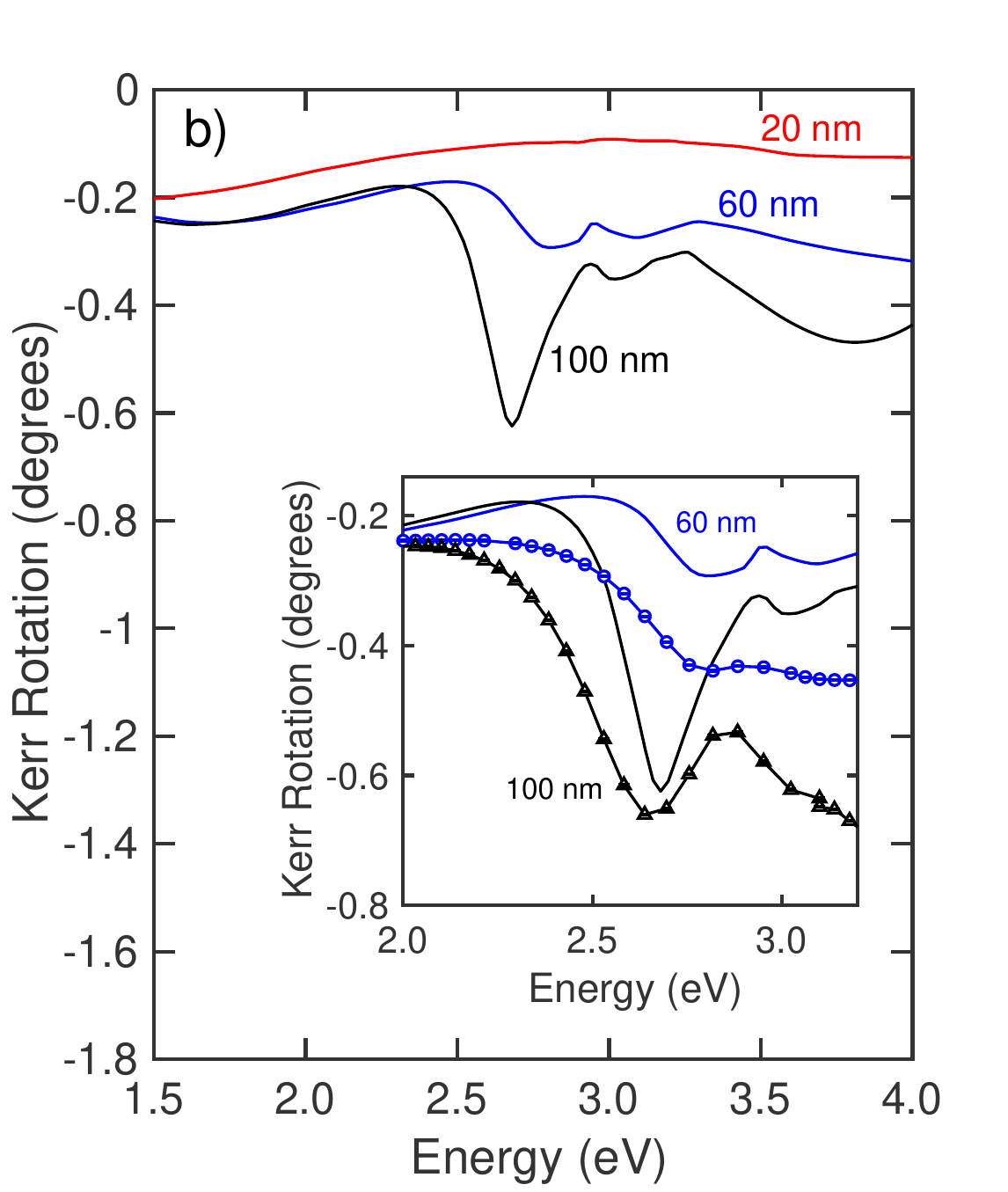}}}$
	\caption{a) Measured spectral polar Kerr rotation for the 100~nm, 60~nm, 20~nm antidot films as well as the 20~nm reference film. The patterned samples show clear changes in the Kerr spectra at different thicknesses. Strong enhancement of the Kerr rotation is obtained from the 60~nm and 100~nm samples. The inset shows a magnified comparison of the 20~nm patterned and reference films in the optical region where the Si becomes highly absorbing. b) Calculated polar Kerr rotation spectra for 20~nm, 60~nm and 100~nm antidot films. The inset shows a comparison with the measured (data points) and calculated (solid lines) data for the 60~nm and 100~nm antidot samples, which shows good agreement between the shape and position of the M-O enhancement.}
	\label{fig3}
\end{figure*}

After discussing the pure plasmonic contribution, we now move to discuss the resulting M-O enhancement. Spectral measurements of the Kerr rotation ($\theta\_{Kerr}$) are shown in Fig.~\ref{fig3}~a). There is a considerable enhancement in the $\theta\_{Kerr}$ when the thickness is increased. This enhancement is over a range of 100~nm, substantially thicker than the penetration depth in Co which is $\approx$~13~nm~\cite{temnov2010} over the measured energy range. The thinnest sample of 20~nm presents a very small signature of SPPs in its reflectivity curve and does not exhibit a significant difference in the Kerr spectrum with respect to its continuous counterpart. For the 60~nm and 100~nm samples we observe maximal $\theta\_{Kerr}$ for a photon energy of approximately $2.81$ eV and $2.69$ eV respectively. These energies correlate well with the observed minima in reflectivity which we established as originating from the excitation of SPPs. The $\theta\_{Kerr}$ enhancement at $2.69$ eV for the 60~nm sample is weaker and appears broader than the counterpart feature at $2.81$~eV for the 100~nm sample.  

A surprising characteristic of the Kerr rotation spectra, shown in Fig.~\ref{fig3}~a), is the behaviour above $3.5$ eV.  For the 60 nm sample, $\theta\_{Kerr}$ continues to increase in magnitude and around $4$ eV is three times higher than the continuous film. Likewise for the 100 nm sample there is a dramatic enhancement of $\theta\_{Kerr}$, peaking at $\approx 3.75$ eV. However, we consider that these features  do not constitute a real $\theta\_{Kerr}$ enhancement and are likely associated with the dramatically reduced reflectivity of the samples in this energy region, as indicated in Fig.~\ref{fig2}~a). Furthermore, as shown in the inset of Fig.~\ref{fig3}~a), both the 20~nm antidot and the reference films show a dip in the same region. We expect that this perceived dramatic enhancement for the 60~nm and 100~nm samples is more likely associated with this weak non-plasmonic enhancement of $\theta\_{Kerr}$ (which is visible even in the 20~nm reference film) and is then exacerbated by the near-zero reflectively of the 60~nm and 100~nm antidot samples in this energy region.

As with the optical reflectivity, the Kerr rotation shows little difference between the continuous 20~nm Co film and the patterned Co film. The same arguments based on the reduction of film conductivity with reducing thickness, along with the fact that at 20~nm the thickness begins to approach $\delta$, can be used to explain the lack of plasmonic activity in the M-O spectra for the 20~nm antidot sample.  

Calculations of the polar Kerr rotation are shown in Fig.~\ref{fig3}~b), capturing well the size and the shape of M-O enhancement. There is excellent agreement between the predicted energies for the SPP enhancement of the polar Kerr rotation signal for the 60~nm and 100~nm samples. This is more clearly demonstrated in the inset of Fig.~\ref{fig3}~b), where it is visible that even the magnitude of the enhancement is in near agreement for the 100~nm sample. Similar to the calculated reflectivity, the calculated Kerr rotation also shows additional features which arise due to the splitting of the plasmonic excitation in different directions as we move away from the normal incidence. These occur at $3.0$~eV and $3.1$~eV for the 100~nm and 60~nm films respectively. Interestingly, in the measured Kerr rotation for the 100~nm sample, there is an indication of a shoulder feature in the spectra at approximately $3.0$~eV, which could be related to this effect.

The very large enhancement at $\approx$ 3.75eV seen in the experiment for the 100~nm antidot sample is not reproduced in the calculations, which further compounds evidence that this is an artifact associated with the near zero reflectivity of the sample in this region.

\section{Conclusions}
In conclusion we have shown a new way to manipulate the M-O response of magnetoplasmonic structures, taking advantage of the thickness of the magnetic layer. We have used patterned Co hexagonal antidot lattices, with different thicknesses, to generate a large enhancement of the polar Kerr rotation as a result of SPP excitation. We reveal that it is not only the in-plane structure which defines the excitation conditions for SPPs, but also that the out-of-plane structure of the magnetic layer plays a crucial role, which can be easily modified by changing the thickness. This is true even for thicknesses well above the penetration depth of the continuous Co film, demonstrating the importance of the enhanced penetration depth accompanied by SPP excitation in antidot structures. The thickness dependence is further confirmed in the related M-O enhancement. We have shown that the thickness dramatically modifies the Kerr rotation enhancement by SPP excitation and consequently provides new routes for tailoring the functionality of patterned structures, where the influence of the thickness on the M-O activity should be taken into account.

\begin{acknowledgments}
The authors acknowledge support from the Knut and Alice Wallenberg Foundation project ``{\it Harnessing light and spins through plasmons at the nanoscale}'' (2015.0060), the Swedish Research Council and the Swedish Foundation for International Cooperation in Research and Higher Education. This work is part of a project which has received funding from the European Union's Horizon 2020 research and innovation programme under grant agreement no. 737093. E. Th. P acknowledges the Deutsche Forschungsgemeinschaft (DFG) through the collaborative research center SFB TRR 173: SPIN+X Project B07 and the Carl Zeiss Foundation. A.G.-M. acknowledges funding from the Spanish Ministry of Economy and Competitiveness through grant MAT2014-58860-P, and from the Comunidad de Madrid through Contract No. S2013/MIT-2740. The authors would like to thank Piotr Patoka for the preparation of the polystyrene bead template. 

\end{acknowledgments}

\bibliographystyle{apsrev4-1}

%

\end{document}